\documentclass{iaus}
\usepackage{graphicx}

\title[short title of paper] 
{Thresholds for the Dust Driven Mass Loss from C-rich AGB Stars}

\author[Mattsson et al.]   
{Lars Mattsson, Rurik Wahlin \& Susanne H\"ofner}%
  
\affiliation{Department of Astronomy \& Space Physics, Uppsala University,
Sweden, \break email: mattsson@astro.uu.se}

\pubyear{2007}
\volume{241}
\pagerange{}
\date{}
\jname{Stellar Populations as Building Blocks of Galaxies}
\editors{A.~Vazdekis \& R.~F.~ Peletier, eds.}
\begin{document}

\maketitle

\begin{abstract}
It is well established that mass loss from AGB stars due to dust driven winds cannot be arbitrarily low. We model 
the mass loss from carbon rich AGB stars using detailed frequency-dependent radiation hydrodynamics including dust 
formation. We present a study of the thresholds for the mass loss rate as a function of stellar parameters based on 
a subset of a larger grid of such models and compare these results to previous theoretical work. Furthermore, we 
demonstrate the impact of the pulsation mechanism and dust formation for the creation of a stellar wind and how it 
affects these thresholds and briefly discuss the consequences for stellar evolution.

\keywords{stars: AGB and post-AGB, atmospheres, mass loss}
\end{abstract} 
\firstsection 
\section{Why a Mass Loss Threshold?}
  As shown by, e.g., Gail \& Sedlmayr (1987), Dominik et al. (1990) and appearent in the detailed models by
  H\"ofner et al. (2003) as well,
  a dust-driven stellar wind cannot be maintained down to arbitrarily small ratio of radiative to graviational 
  acceleration $\Gamma_{\rm d}$. For a "polytropic wind" one may derive an analytical expression for the terminal
  wind velocity,
  \begin{equation}
  \label{terminal} 
  v_\infty^2 \approx {1\over 2}\Delta v^2_{\rm p} + \left({2\over \gamma -1}\right)\,\bar{c}_s^2(R_{\rm in}) 
  + \bar{v}^2_{\rm esc}(R_{\rm c}) \left[\bar{\Gamma}_{\mathrm{d}}-{R_{\rm c}\over R_{\rm in}}\right],
  \end{equation}
  where $\gamma$ is the polytropic index, $R_{\rm c}$ is the characteristic radius at which dust starts to condense, 
  $ \bar{v}^2_{\rm esc}(R_{\rm c})$ is the average escape velocity at $R_{\rm c}$, $\bar{c}_s^2(R_{\rm in})$ is the 
  average sound speed at the inner boundary of the model (located at $R_{\rm in}\sim R_\star$) and $\Delta v_{\rm p}$
  is the "piston amplitude", i.e. the strength of the pulsations (Mattsson 2006). The equation
  above captures the general trend of $v_\infty$ with $\Gamma_{\rm d}$ and predicts a threshold at 
  $\Gamma_{\rm d}\approx0.8$ for reasonable values of the model parameters (see Fig. 1, left panel).

\begin{figure}
  \includegraphics[width=6.5cm]{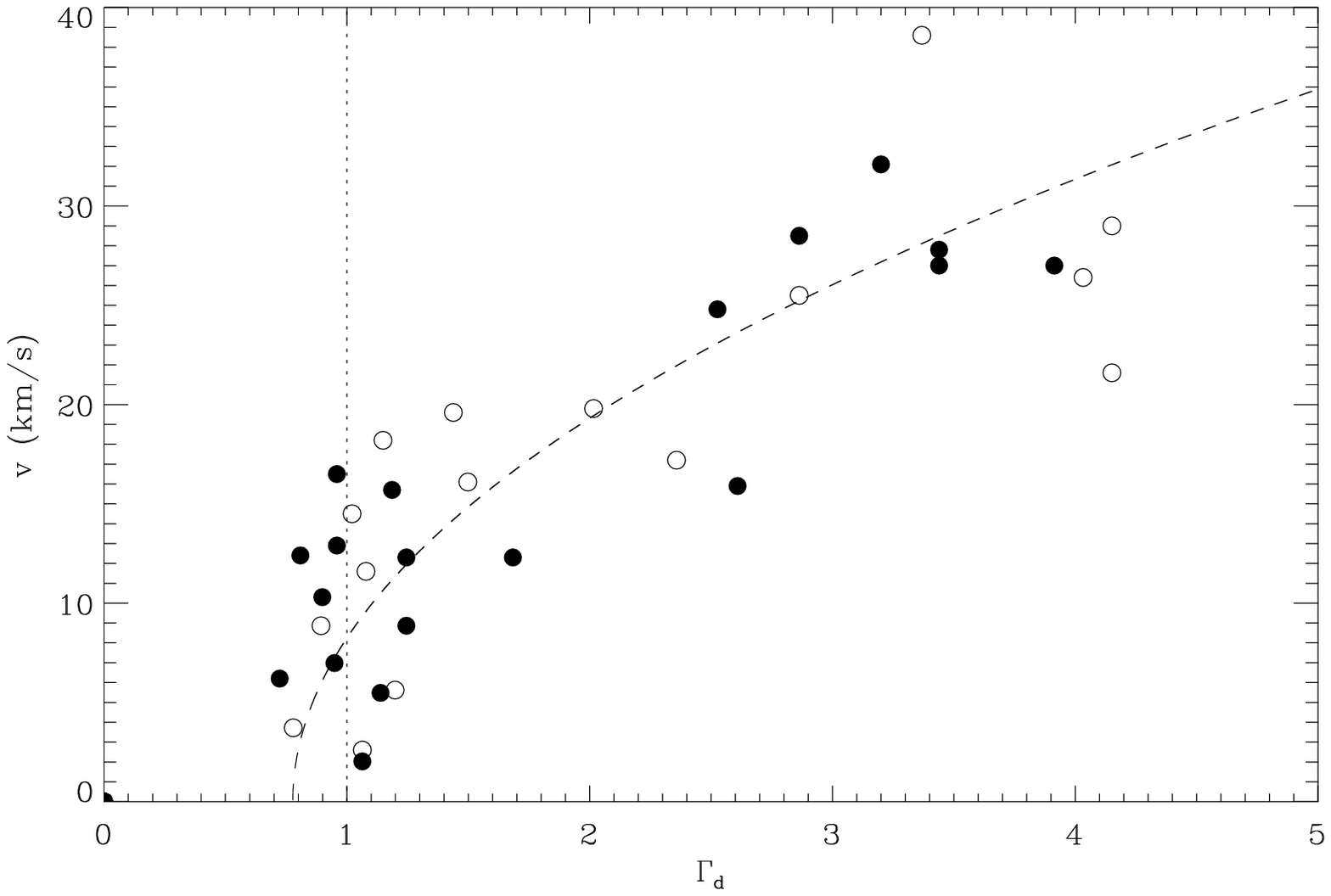}
  \includegraphics[width=6.5cm]{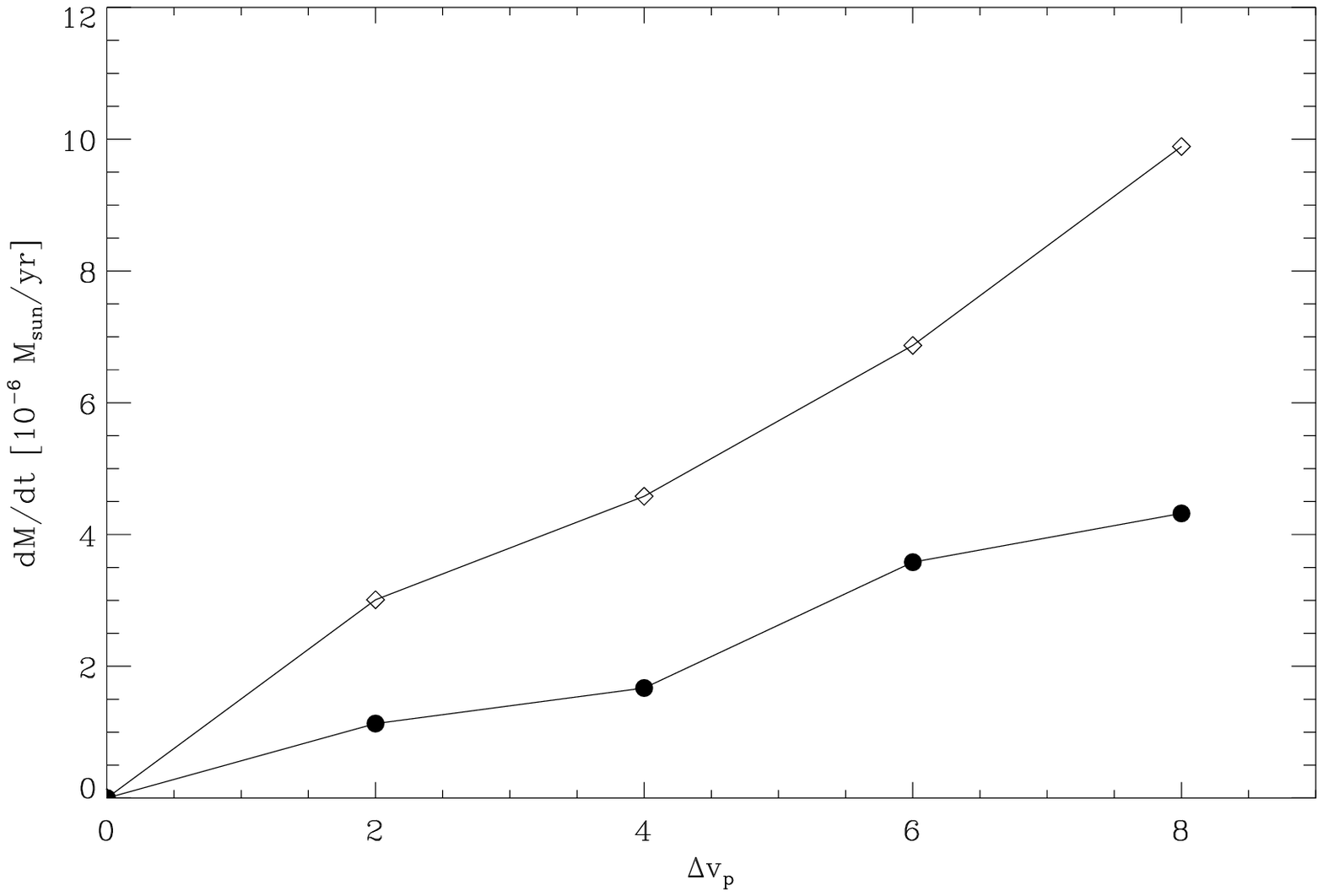}
\caption{Left: Wind velocity (in km s$^{-1}$) as function of the acceleration parameter $\Gamma_{\rm d}$ for all models 
of the sub-grid calculated with $M_\star = 1M_\odot$, $Z = Z_\odot$ and $\Delta v_{\rm p} = 4.0$ km s$^{-1}$. Black dots 
represent models with $\Delta v_{\rm p} = 4.0$ km s$^{-1}$ and circles represent models with $\Delta v_{\rm p} = 6.0$ km s$^{-1}$. 
The dashed curve shows an analytical model with $\gamma = 7/6$, $M_\star = 1M_\odot$, $R_\star = 3.5\cdot 10^{13}$ cm, 
$R_{\rm in}(0) = 0.9 R_\star$, $R_{\rm c} = 2.5 R_\star$ and $c_s(R_{\rm in}) = 7.0$ kms$^{-1}$. Right: The mass loss rate 
as a function of the piston amplitude. Black dots represents the case where $L_\star = 7100 L_\odot$ and diamonds represents 
$L_\star = 10000 L_\odot$.} 
\end{figure}

  We have used our RHD code for dynamic stellar atmospheres of carbon-rich AGB stars (described in H\"ofner et al. 2003, 
  Mattsson 2006), including frequency-dependent radiative transfer and dust formation, to explore the relations 
  between basic stellar parameters and a dust-driven stellar wind.  Here we present results from the computation of 
  a grid of wind models at solar metallicity. An associated library of dynamic spectra is under development (see the
  poster by Wahlin et al.).

  A mass loss threshold appears as one would expect from Eq. (\ref{terminal}) and we find that below a critical ${\rm C/O}$ 
  and/or above a critical $T_{\rm eff}$ no dust driven wind can be formed.  All other stellar parameters were 
  kept fixed in these models. We also see a rather strong dependence on ${\rm C/O}$ for both the wind velocity and the mass 
  loss rate, which is quite interesting in comparison with previous studies of this kind. Arndt et al. (1997) as well as 
  Wachter et al. (2002) find a weak dependence on ${\rm C/O}$, which stands in sharp contrast to the results presented here. 
  However, our findings here (as well as in H\"ofner et al. 2003) are, qualitatively speaking, hardly a new discovery. 
  H\"ofner and Dorfi (1997) and Winters et al. (2000) have already pointed out the strong C/O-dependence, especially in 
  the critical wind regime, although this has not been widely recognised. Furthermore, there is a linear dependence of the 
  mass loss rate on the piston amplitude, i.e. $\dot{M} \propto \Delta v_{\rm p}$ (see Fig. 1, right panel). The trend is 
  strong enough to make $\Delta v_{\rm p}$ significant in parametric prescriptions of the mass loss rate.   

\section{Conclusions}
The results from our new detailed grid of wind models at solar metallicity suggests that C-stars with strong winds 
may actually be a rare species. How would this affect models of stellar evolution, nucleosynthesis and, consequently, 
models of chemical evolution of galaxies? We want to make the following points:
\begin{itemize}
\item The strength of the pulsations and the C/O-ratio are {\it not} redundant parameters in a mass loss prescription.
\item It may be dangerous to use parametric mass loss formulae including too few stellar parameters and extrapolate
beyond the range of stellar parameters used obtain the formula.
\item The mass loss rate depends strongly on the efficiency of dust formation, which cannot simply be parameterised in
terms of the basic stellar parameters: mass, luminosity and temperature, only.
\item {\bf There exists a threshold for dust-driven winds which has previously been neglected in mass loss
prescriptions and thus not included in models of stellar evolution!}
\end{itemize}


\end{document}